\documentclass[pdflatex,sn-mathphys-num]{sn-jnl}

\usepackage[T1]{fontenc}
\usepackage[utf8]{inputenc}

\usepackage{graphicx}%
\usepackage{multirow}%
\usepackage{amsmath,amssymb,amsfonts}%
\usepackage{amsthm}%
\usepackage{mathrsfs}%
\usepackage[title]{appendix}%
\usepackage{xcolor}%
\usepackage{textcomp}%
\usepackage{manyfoot}%
\usepackage{booktabs}%
\usepackage{algorithm}%
\usepackage{algorithmicx}%
\usepackage{algpseudocode}%
\usepackage{listings}%
\usepackage{tabularx}

\theoremstyle{thmstyleone}%

\theoremstyle{thmstyletwo}%

\theoremstyle{thmstylethree}%

\begin{document}

\title[AI Researchers’ Views on Automating AI R&D and Intelligence Explosions]{AI Researchers’ Perspectives on Automating AI R\&D and Intelligence Explosions}


\author{\fnm{Severin} \sur{Field}}\email{severin.field@louisville.edu}
\author[]{Raymond Douglas}
\author[]{David Krueger}
\affil{\orgdiv{ML Alignment and Theory Scholars Program,
\orgaddress{\city{Berkeley}, \postcode{94701}, \state{CA}, \country{USA}}}}


\abstract{Many leading AI researchers expect AI development to exceed the transformative impact of all previous technological revolutions. This belief is based on the idea that AI will be able to automate the process of AI research itself, leading to a positive feedback loop\cite{wiseman_how_2025}. In August and September of 2025, we interviewed 25 leading researchers from frontier AI labs and academia, including participants from Google DeepMind, OpenAI, Anthropic, Meta, UC Berkeley, Princeton, and Stanford to understand researcher perspectives on these scenarios. Though AI systems have not yet been able to recursively improve, 20 of the 25 researchers interviewed identified automating AI research as one of the most severe and urgent AI risks. Participants converged on predictions that AI agents will become more capable at coding, math and eventually AI development, gradually transitioning from `assistants' or `tools' to `autonomous AI developers,' after which point, predictions diverge. While researchers agreed upon the possibility of recursive improvement, they disagreed on basic questions of timelines or appropriate governance mechanisms. For example, an epistemic divide emerged between frontier lab researchers and academic researchers, the latter of which expressed more skepticism about explosive growth scenarios. Additionally, 17/25 participants expected AI systems with advanced coding or R\&D capabilities to be increasingly reserved for internal use at AI companies or governments, unseen by the public. The concerns most frequently raised by participants were power concentration and AI-augmented progress accelerating behind closed doors, and they offered different suggestions for mitigation strategies or explanations for why certain mitigations might turn out worse than the problem they seek to solve. Participants were split as to whether setting regulatory ``red lines" was a good idea, though almost all favored transparency-based mitigations.
}


\maketitle

\section{Introduction}\label{sec1}
An `intelligence explosion' is a hypothetical scenario introduced by I.J. Good in 1966\cite{alt_speculations_1966}, in which AI systems become capable of recursively improving their own capabilities (designing AI successors that can design AI even better). Yudkowsky \cite{yudkowsky_intelligence_nodate} defines an intelligence explosion in terms of the return on cognitive investment, when investing cognitive resources in improving cognition accelerates the returns from those investments. Recursive improvement need not look like conventional AI R\&D, future AI systems might be able to directly plan and program recursive improvements, or improve by improving their physical hardware. However, a critical capability milestone may occur when AI systems can do the work of today's AI researchers, or AI R\&D, at which point, such an AI could improve the development of it's successor. 

In August of 2025, a version of OpenAI's GPT-5 model won a gold medal at the International Math Olympiad\cite{castelvecchi_deepmind_2025}. According to OpenAI's Chief Scientist Jakub Pachocki, one of OpenAI's main priorities is to ``automate scientific discovery'' with a plan to build automated researchers improving AI capabilities further\cite{openai_agi_2025}. The recursive improvement discussion is also taking place in academic venues. For example, ICML 2026, one of the largest and most prestigious machine learning conferences, is hosting a workshop on ``AI with Recursive Self Improvement." At the same time, many researchers reject the notion of an `intelligence explosion'\cite{chollet_implausibility_2018} and the field of AI has a reputation for failing to deliver on massive promises. 

While leading AI companies race toward systems capable of automating AI research, a development many consider the field's most urgent risk, AI researchers remain divided on the risks, timelines and possibility of recursive improvement and intelligence explosions.

We conducted a series of semi-structured interviews to understand the perspectives of the leading researchers working on AI systems for AI R\&D. We refer to such systems as ASARA (AI Systems for AI R\&D Automation). This work is relevant to understand: 
\begin{enumerate}
    \item how researchers envision AI systems automating AI research and which key milestones exist
    \item where researchers disagree over intelligence explosion scenarios
    \item what specific risks arise from AI-automated R\&D and how these compare to other AI-related concerns
    \item expert views on risk mitigations, governance, and red lines
\end{enumerate}

We interviewed 25 leading researchers from frontier AI companies and academia to document their views on these scenarios and capture the reasoning for researchers' views. The participant selection strategy prioritized a diversity of perspectives while maintaining rigorous inclusion criteria (AI researchers at frontier labs and researchers with published work on the topic of recursive improvement at a top venue; the methodology section outlines selection criteria).

``AI self improvement'' is already widely considered a red line by many decision makers\cite{hausenloy_red_2025,zoumpalova_global_2025}. For instance, the IDIAS Beijing Consensus declares ``No AI system should copy or improve itself without explicit human approval''\cite{noauthor_idais-beijing_nodate}. Yet my findings reveal the skepticism around risk mitigation and have implications for improving multilateral agreements. Table \ref{tab:definitions} shows the key definitions used in this study.
\begin{table}[h]
\caption{Key terminology and definitions as used during interviews. When the terms first occurred during interviews, participants were given these definitions.}\label{tab:definitions}
\begin{tabularx}{\textwidth}{lX}
\toprule
\textbf{Term} & \textbf{Definition} \\
\midrule
Intelligence Explosion & 
AI developing more advanced AI, which would itself develop even more advanced AI, faster and better than the last time, resulting in exponential growth in AI capabilities. Past a certain point, the process is self-sustaining and no longer requires human intervention. \\
\addlinespace
ASARA & 
AI Systems for AI R\&D Automation. We define ASARA as ``AI systems which are able to meaningfully contribute to the development of frontier AI models.'' This acronym comes from Eth\cite{eth_will_nodate}, which defines ASARA as ``being able to substitute for any remote R\&D workers at companies advancing the state of the art for AI.'' The interviews use a broader definition to capture the scenario the participants might envision without leading them.  \\
\addlinespace
Frontier AI Company & 
An AI company at the frontier of LLM capabilities. We use ``frontier lab" interchangeably because this is often how they self-identify. However, in our work, both terms refer exclusively to: xAI, Meta, Google DeepMind, OpenAI and Anthropic. \\
\addlinespace
Red Lines & 
A limit that, if crossed, should trigger a big response by AI developers or governments. \\
\bottomrule
\end{tabularx}
\end{table}
\subsection{Related Work}

The prospect of artificial intelligence systems capable of improving itself has evolved from philosophical speculation\cite{alt_speculations_1966} to an active research priority of AI researchers\cite{openai_agi_2025}. Early theoretical work from Bostrom\cite{bostrom_superintelligence_2014} and Yudkowsky\cite{yudkowsky_intelligence_nodate} argues that such recursive self-improvement could lead to an intelligence explosion with transformative consequences. Critics including Chollet\cite{chollet_implausibility_2018} argue that fundamental constraints on recursive improvement, or diminishing returns, prevent the rapid emergence of systems beyond human comprehension and control.

\textbf{Benchmarks Measuring Progress:} Benchmarks like PaperBench\cite{starace_paperbench_2025}, REBench\cite{wijk_re-bench_2025}, and MLEBench\cite{chan_mle-bench_2025} measure AI's ability to conduct machine learning research, while systems like R\&D agents demonstrate growing automation capabilities\cite{yang_rd-agent_2025}. In order to address the benchmark saturation problem, METR has introduced an unbounded task horizon benchmark measuring AI performance in terms of the length of tasks AI agents can complete\cite{kwa_measuring_2025}. METR has also studied ``uplift'': productivity improvement gained by augmenting humans with AI tools\cite{becker_measuring_2025}.


\textbf{Interviews of Researchers:} Two recent interview studies\cite{owen_interviewing_2024,leibowich_could_nodate} have explored the perspectives of researchers on AI automation. Owen\cite{owen_interviewing_2024} interviewed 8 researchers focusing on characterizing the difference between AI R\&D work tasks and timelines for when they'd be reached. Leibowich et al.\cite{leibowich_could_nodate} interviewed 5 researchers with a hypothetical scenario about full automation of AI R\&D, focusing on bottlenecks and where AI cognitive labor would accelerate progress. However, these studies primarily focused on bottlenecks and capability forecasts, while this study seeks to understand risks, mitigations, and what various organizations think about ASARA.

\textbf{Forecasting:} Davidson et al.\cite{davidson_how_nodate} have published concrete mathematical models of an intelligence explosion, and concrete scenario forecasts such as AI 2027 by Kokotajlo et al.\cite{kokotajlo_ai_nodate} have gained traction among AI researchers.


\section{Methodology}\label{sec:methodology}

\subsection{Participant Recruitment}\label{subsec:recruitment}

In total, we invited 182 researchers to participate, 25 of whom agreed to be interviewed. Participants came from three recruiting approaches to capture different perspectives on AI self-improvement dynamics:

\begin{enumerate}
    \item \textbf{Literature-Based Recruitment (7 participants)}: We conducted Google Scholar searches with the following search terms: ``recursive self-improvement," ``intelligence explosion," ``AI automating ML research," ``AI R\&D Automation." Then we contacted authors from papers that introduced novel technologies for AI-assisted research or offered critiques or defense of intelligence explosion scenarios.
    \item \textbf{Conference Workshop Recruitment (8 participants):} We identified participants from relevant workshops at NeurIPS and ICLR 2024. At ICLR, these included: the Third Deep Learning for Code (DL4C) Workshop, Towards Agentic AI for Science Workshop, and Self-Improving Foundation Models Without Human Supervision. We also recruited participants from accepted papers at the NeurIPS Workshop on Scalable Continual Learning for Lifelong Foundation Models.
    \item \textbf{Network Based Recruitment and Snowball Sampling (10 participants):} Initial participants were asked to recommend others who ``might have thought deeply about this, might have a different angle, or disagree with [them]," as well as through professional connections established during the ML Alignment \& Theory Scholars (MATS) program.
\end{enumerate}

\subsection{Interview Protocol}\label{subsec:protocol}

Interviews were conducted in August and September of 2025 and lasted 40--60 minutes each. At the beginning of each interview, we described the goals of the study and the practices used to ensure participants' anonymity. We obtained verbal consent to participate before questions. 

The interview protocol was organized into three sections. The first section explored whether they expected AI systems to automate AI research, what it might look like, and the type of trajectory they expect. The second section addressed organizational dynamics, including deployment decisions and how various actors could and should respond. The third section focused on risks and governance, including what mitigations and regulatory red lines could be put in place. The interview protocol can be found in Appendix \ref{app:interview-protocol}. A full list of (anonymized) participants can be seen in Table \ref{tab:participants}.

The interviews were recorded on Google Meets and transcribed using OpenAI whisper or Otter.AI. Participants were made anonymous by default, informed they were free to skip questions, and given the opportunity to review their quotes before publication. The full Participant Information Sheet and list of interview questions can be found in the Appendix.

\begin{table}[h]
\caption{Overview of study participants, including role descriptions and how they are grouped during qualitative coding. }\label{tab:participants}
\centering
\begin{tabular}{lll}
\toprule
\textbf{ID} & \textbf{Affiliation Type} & \textbf{Role Description} \\
\midrule
1  & Frontier AI Company& Research Scientist \\
2  & Frontier AI Company& Research Scientist \\
3  & Frontier AI Company& Research Scientist \\
4  & Frontier AI Company& Research Scientist \\
5  & Frontier AI Company& Research Scientist \\
6  & Frontier AI Company& Research Scientist \\
7  & Frontier AI Company& Research Scientist \\
\midrule
8  & Former Frontier AI Company& Research Scientist \\
9  & Former Frontier AI Company& Research Scientist\\
10 & Former Frontier AI Company& Research Scientist\\
11 & Former Frontier AI Company& Research Scientist\\
\midrule
12 & Academia & Professor\\
13 & Academia & Professor \\
14 & Academia & Professor \\
15 & Academia & Postdoctoral Fellow \\
16 & Academia & PhD Student\\
17 & Academia & PhD Student\\
18 & Academia & PhD Student \\
19 & Academia & PhD Student \\
20 & Academia & PhD Student (Adjacent Scientific Field) \\
\midrule
21 & Industry & Research Scientist at a Big Tech Company\\
22 & Industry & Research Scientist at a Big Tech Company\\
23 & Industry & Startup Founder \\
\midrule
24 & Nonprofit & Forecasting Researcher \\
25 & Nonprofit & Research Scientist \\
\bottomrule
\end{tabular}
\end{table}

\subsection{Analysis}\label{subsec:analysis}

We transcribed the interviews with a private instance of OpenAI Whisper or Otter.AI and then de-identified interviews following Saunders et al.\cite{saunders_anonymising_nodate}, where participant names were replaced with pseudonyms, and institutional affiliations were generalized. 

Codes were developed inductively after reading the transcripts. When similar concepts emerged across multiple interviews (e.g. participants mentioning a particular constraint), these were tagged as codes. Related codes were then grouped into higher order themes. Some research questions included: what the technology might look like, clarity of the trajectory, mechanisms underlying risks, participants' beliefs about red lines, beliefs about mitigations, organizational realities, and beliefs about internal deployments.

Inductive coding was done by the first author after reading the transcripts and conducting interviews. For categorical variables presented in Figures 1-4, we employed AI-assisted coding. The AI system (Claude) was provided with anonymized transcripts one at a time in separate instances, using structured prompts to identify participant positions (allowing an ``unsure" option). An example prompt can be found in Appendix \ref{app:ai_prompts}.

\section{Results}\label{sec2}
This section presents our findings in five parts. Section \ref{subsec:asara-vision} examines what participants expect ASARA will look like in the coming years. Next, Section \ref{subsec:publicvsinternal} explores participants' expectations about public and internal deployments; notably, many participants expected frontier labs to keep their most capable models internal\footnote{When participants referred to `internal' deployment, they generally meant withholding from public or commercial release, not necessarily from government agencies.}, which informed later questions about governance and observability. Next, Section \ref{subsec:trajectory} addresses participants' beliefs about the trajectory of ASARA and where constraints may lie. Finally, Section~\ref{subsec:risk-perception} presents participants' risk perceptions, and Section~\ref{subsec:red_lines} discusses views on red lines and alternative governance mechanisms.

\subsection{What Participants Imagine ASARA Looking Like}\label{subsec:asara-vision}

Participants converged on a pathway to ASARA where AI systems gradually augment programming and then become more autonomous, eventually capable of scientific discovery. 17 participants described a progression similar to this, with Participant 1 capturing the consensus view: ``We first get a model that's really good at coding. A model that's really good at coding is almost already very good at ML R\&D, maybe needs a little bit of actual work, but then you get superhuman ML R\&D.'' 

This progression can be conceptualized in three stages:
\begin{enumerate}
\item \textbf{Research speedup stage}: AI systems ``raise the floor'' of all researchers' abilities, speeds up research, but remain bottlenecked by human oversight. Participants commonly described tools that multiplied their productivity, for example, Participant 7 described expecting systems ``like Claude Code (Anthropic's AI coding assistant\footnote{https://claude.com/product/claude-code}) but [that] add 5x to my [coding] speed." 
\item \textbf{Collaboration}: AI autonomously handles research sub-tasks while humans maintain high-level direction. Participant 7 described their vision of this stage as, ``Cursor but like three to five times my speed."
\item \textbf{Full loop automation stage}: AI systems independently execute complete research cycles. At this stage, Participant 12 said, ``the human will become the bottleneck, the companies will try to remove the humans, by all means." 
\end{enumerate}

Participants used terms including ``intelligence explosion," ``takeoff," ``singularity," ``recursive improvement" relatively similarly, and all but two (P20 and P15, who did not like the term `intelligence explosion') were comfortable discussing, or at least speculating, about an intelligence explosion as if it were the natural consequence of automating AI research, despite large differences such as when it would take place or how quickly.

Participant 1 said, ``I do think that though it is mocked, the term singularity is good. I view it as an event horizon. I just, like, can't see what it looks like to have built something that's self-improving so quickly." 

A subset of participants (occurring in 4 transcripts) introduced the concept of \textbf{task horizon progression} as a key metric for tracking ASARA development. They described a trajectory from current systems capable of $\sim$1--2 hour software engineering tasks to future systems handling 40+ hour research projects autonomously. 

Many also expected \textbf{internal deployments}, where frontier capabilities remain in labs as opposed to being publicly released.

\subsection{Public versus Internal Deployment Expectations}\label{subsec:publicvsinternal}



\begin{figure}[h]
\centering 
\includegraphics[width=\textwidth]{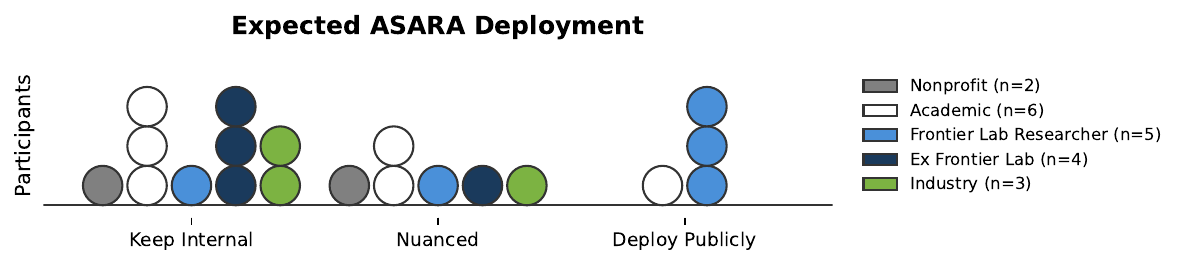}
\caption{Participant predictions on whether AI companies will publicly deploy models capable of meaningfully accelerating AI research, grouped by affiliation type. The figure shows a categorical dot plot with participants grouped into ``Expects Internal,'' ``Nuanced,'' and ``Expects Public'' categories. Each dot is a participant; a full participant table can be found in Table \ref{tab:participants}. Participants with an unclear opinion are omitted from the figure. Participants who expect public deployment cite economic pressures, funding needs, competitive dynamics, and responsible decision-making. Those expecting internal deployment believe ASARA is too valuable to share and offers higher returns when used for internal R\&D rather than API deployment. The ``Nuanced'' category includes participants who believe deployment decisions depend on government intervention and inter-lab differences in policies and culture.}
\label{fig:deployment_predictions}
\end{figure}

In the past, AI companies have kept developments internal. For instance, OpenAI reported spending six months on safety research, risk assessment, and iteration before releasing GPT-4\cite{openai_gpt-4_2024}. As seen in Figure \ref{fig:deployment_predictions}, half of the 20 participants who clearly addressed this question expected AI labs to keep AI research-capable models internal  and only 20\% expected them to deploy them publicly.

Participants suggested that keeping models internal could offer it's developers serious advantages, possibly by accelerating their R\&D efforts. Participant 25 said, ``internal-only deployments might happen, and that is a big risk factor, because there's just less information.'' On the other hand, participants identified arguments and pressures that would promote diffusion of AI capabilities, including economic pressure to commercialize AI capabilities, insiders leaking milestones, and a culture of boasting about capabilities. Participants who believed internal deployments were more likely to express concerns about oversight and lack of external scrutiny.

However, framing development as a binary choice may be too simplistic. Six participants gave nuanced perspectives, three suggested that outcomes would vary by company. For instance, Meta's open source culture differs sharply from OpenAI's more closed approach. AI companies could also sell weaker versions of their models while retaining full capabilities internally. As Participant 18 explained, ``They'll train a base model, then they won't release that model, not only because it's not economical but also because it risks distillation, but they will distill it themselves to their cheaper models that they do inference on."

Despite this complexity, clear patterns emerged in participants' reasoning. The following sections present the most common arguments for internal versus public deployment.

\subsubsection{Arguments That Labs Will Keep ASARA Models Internal}

Many participants expected AI companies reaching ASARA capabilities to avoid releasing their technology to the public. The most common codes emerging for this were:

\begin{enumerate}
\item \textbf{Preserving Competitive Advantage} (occurred in 12 transcripts): Participants suggested that deploying models could accelerate their competitors. This code captures the economic logic that providing competitors with capability-enhancing tools harms a developer's market position.

\item \textbf{Limited compute} (occurred in 10 transcripts): Researchers described a trade-off between using limited computational resources for internal R\&D acceleration versus external service provision. Using limited computational resources to serve the public AI access via API, for example, means those resources cannot be used for AI R\&D. Participant 10 argues that because AI research is such a highly valued economic niche, AI companies will likely find more value keeping the models internal. For instance, ``\$100,000 in compute resources might be worth \$1,000,000 [in AI researcher salary].''

\item \textbf{Security / Avoiding Diffusion} (occurred in 6 transcripts): Participants expect frontier AI companies to restrict deployment due to capability diffusion concerns. One potential explanation for this is not trusting others to use the technology responsibly. For example, public deployments or open-source deployments allow capabilities to be reverse-engineered, distilled, or lose value. ``They'll train a base model, and then they won't release that model, not only because it's uneconomical but also because it risks distillation'' (P18). 
\end{enumerate}

\subsubsection{Arguments That Labs Will Deploy ASARA Models Publicly}

\begin{enumerate}
\item \textbf{Financial pressures} (10 transcripts): The most prevalent rationale for public deployment centered around AI companies' dependence on investor funding and revenue generation. Companies expending capital on R\&D face existential pressure to deploy products, which might override strategic internal deployment considerations. Similarly, some participants suggest that no developer can afford to withhold capabilities while competitors deploy their systems. One participant articulated this by arguing, ``I think people overrate the chances that they will be kept secret and internal... The labs need to and want to win against each other. Holding a model back is expensive... I think a lot of researchers live in this world where their main concern is super intelligence, but labs, you know, they need to raise another \$20 billion every year... the key thing for that is managing investor expectations and showing revenue growth and so forth.'' (P23, Startup Founder)

\item \textbf{Government and regulatory intervention} (7 transcripts): Researchers frequently suggested that government actors might step in to prevent internalization of ASARA capabilities. One participant stated: ``The US government would intervene and the FBI would raid them'' (P4), while others described more gradual regulatory processes involving transparency requirements, evaluations, and mandatory access for government agencies.

\item \textbf{Business model and culture} (6 transcripts): Some AI leaders deeply value openness, and this value may reflect in decisions. This code reflects researchers' recognition that different companies face varying incentives based on their revenue structures and corporate contexts. For example, participants distinguished between companies that depend heavily on API revenue versus those with alternative funding sources or objectives. Research labs integrated within larger tech companies might have more flexibility to keep capabilities internal than small companies or nonprofits. Other participants suggested that culture at various institutions differs, with some more likely to favor openness.

\item \textbf{Specialized deployment value} (4 transcripts): This code represents researchers' views that external deployment creates value that exceeds internal-only utilization. Labs lack domain expertise across all potential applications, which creates pressure to sell AI capabilities to high-paying customers who might find niche uses for such capable models. Additionally, researchers described scenarios where distilled or capability-restricted versions could be deployed while maintaining the most advanced systems internally, allowing labs to capture deployment value without sacrificing competitive advantages. As one researcher explained, ``I think there's just vast incentives for them to trade with the rest of the world and uplift other people rather than trying to automate it all the way themselves, when [they] have incredibly less domain expertise in every other field apart from AI.'' (P4, Frontier Lab Researcher)

\item \textbf{Knowledge diffusion inevitability} (3 transcripts): Several researchers expressed skepticism about labs' ability to maintain capability advantages even if they attempt internal retention, citing information leakage via employee mobility, espionage, parallel discovery, and the difficulty of maintaining secrets.
\end{enumerate}

\subsection{Constraints and Milestones} \label{subsec:trajectory}
After establishing what a participant expects ASARA to look like and how it could be deployed, the interviews discussed the path to achieving those capabilities. Participants were asked why they did or did not expect a clear, continuous trajectory towards automating research and development, where the constraints lie, what milestones we might notice, and what evidence confirming or rejecting the participant's prediction might look like. Often, after the participants prediction, we asked, ``what evidence or observations would change your mind?" This section addresses three major themes: (1) disagreement over a clear vs. obstacle-laden path, (2) what constraints may be binding, and (3) proposed observable milestones. 

\subsubsection{Trajectory Clarity}\label{subsec:trajectory_clarity}

\begin{figure}[h]
\centering
\includegraphics[width=\textwidth]{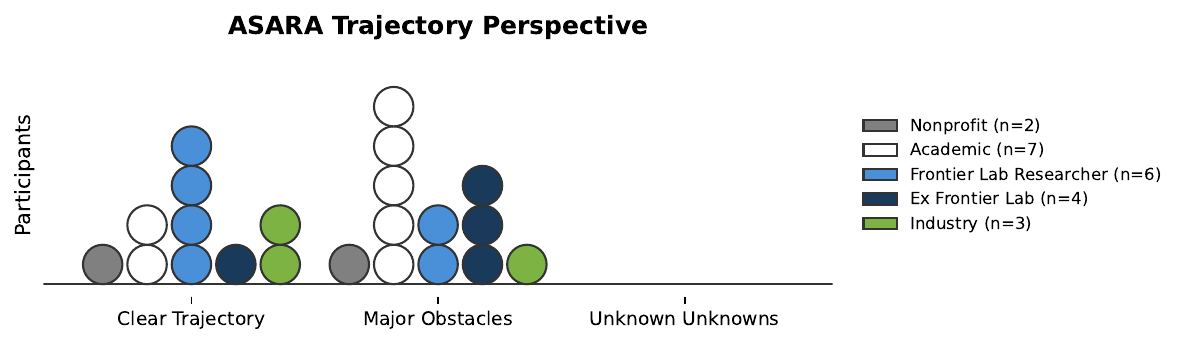}
\caption{Participant views on the clarity of the trajectory toward ASARA, grouped by affiliation. The figure shows participants distributed along a spectrum from ``Clear Path'' through ``Major Obstacles'' to ``Unknown/Unknowns.'' Frontier lab researchers predominantly cluster toward viewing the path as clear, while academic researchers are more distributed across the spectrum, with several identifying major obstacles. Ex-frontier lab researchers show mixed views, and nonprofit/industry participants fall primarily in the middle range (major obstacles; no one fell into the final category `unknown unknown' using the method in Appendix \ref{app:ai_prompts}).}
\label{fig:trajectory_clarity}
\end{figure}

A disagreement between the type of progress expected emerged. Participants disagreed about whether there already exists a relatively clear, continuous trajectory towards automating AI research or if there are large obstacles to overcome, described by Participant 7 as a ``quantum leap." Figure \ref{fig:trajectory_clarity} shows the distribution of participant views about the trajectory towards ASARA.

The divergence in views often corresponded with institutional affiliation: researchers at frontier companies predominantly viewed the path as technically feasible through incremental improvements, while academic researchers more frequently identified fundamental obstacles requiring conceptual breakthroughs. Furthermore, qualitative data from participants suggests that their colleagues with similar affiliations tend to think similarly. 

One professor articulates what the `clear trajectory' perspective looks like, saying, ``[I expect something like] auto ML, optimizing model performance by basically using AI to search through the space of\ldots decisions that affect the model, from hyper parameters to, well, pretty much any kind of parameter which is not a parameter of the model itself, but that can be varied during the training or post training or inference. It's like essentially a humongous search in that hyper parameter space\ldots with the proper amount of compute and conscious effort, which can keep improving and [begin] automating what AI researchers do now\ldots Continuously optimizing and having feedback loop, basically for their own performance\ldots I mean, I don't see any obstacle to automating them (doing the work of a PhD student).'' (P13, Professor)

On the other hand, a PhD student argued: ``Hard ideation as in ideas that actually touch on groundbreaking paradigm shifting ideas\ldots I don't think current language models would be able to do that. First of all, it's very difficult to evaluate. A lot of times, there's no way you can predict this is a paradigm shifting idea before it happens\ldots it also just requires a deeper level of intelligence than just brainstorming simple, low-hanging-fruit kind of ideas.'' (P17, PhD Student)


\subsubsection{Binding Constraints}\label{subsubsec:bindingconstraints}
Sixteen of the transcripts include potential constraints that could limit or fundamentally alter the trajectory towards autonomous AI research. These include:

\begin{itemize}
\item \textbf{Compute Limiting Factor} (occurred in 11 interviews): Some viewed computational limits as engineering challenges amenable to scaling, while others saw physical limits as potential barriers to explosive growth scenarios. The accessibility of compute also raised concerns about market concentration.

\item \textbf{Data Limiting Factor} (occurred in 5 interviews): Multiple participants considered data quality or quantity to be a binding constraint in building ML systems capable of meaningfully contributing to research. For instance, Participant 3 said, ``I don't see a limit necessarily other than maybe data\ldots [data] determines the ceiling of where self-improvement can get you.''

\item \textbf{Bootstrapping Level}: Participants disagreed on the level of capability required for autonomous improvement. All but one acknowledged the possibility, while skeptics argued fundamental breakthroughs would be needed to escape local optima and sustain improvement without human oversight. Participant 1 articulated a specific threshold: ``I do think there is a bootstrapping level. There are some key thresholds you need to be able to hit. Can the model evaluate its own answers? This is a toy version of the problem. Can it rank its own answers? Until relatively recently, the models just could not do that." They continued, arguing that once this threshold is crossed, ``one massive binding constraint, which is human time, which currently is a very key constraint on progress, will drop out."
\end{itemize}

15 of the participants also described a capability distinction that may prove challenging. Despite using different language, both optimists and skeptics distinguished between research ideation and execution capabilities, often expecting the former to be more difficult. Ideation was sometimes discussed as `creativity' or `research taste.' While many gravitated towards `creativity' as something missing, two participants identified the bottleneck in ideation as `validation.' They reasoned that generating large quantities of ideas poses relatively little challenge for AI systems; the more important challenge would lie in discriminating bad ideas from good ones. The forecasting researcher articulated this distinction clearly: ``I think it makes sense to kind of separate out AI R\&D into kind of like two clusters of skills... experiment selection or like more generally maybe like research taste... Then the other is like experiment implementation.'' (P24, Nonprofit) Participant 17 elaborated on why the former may prove more difficult, arguing research ideas as having a `long tail,' where most are mediocre while a minority are highly impactful. They argued that capturing the valuable skill of evaluating ideas will be incredibly difficult, as expert humans with decades of experience struggle with this task, and ML models tend to learn the mode of their data rather than the exceptional cases.


\subsubsection{Observable Milestones and Evidence}
Participants were asked what milestones they expected to see, or evidence that would considerably change their view. Twenty-two of the transcripts contain such discussions. For a comprehensive list of milestones that the participants suggested would update them dramatically on this question, please see the appendix \ref{app:milestones}. Some identified key indicators include:

\begin{itemize}
\item \textbf{Coding Capabilities} (occurred 8 times): Coding capabilities emerged as the most concrete near-term indicator. Participant 12 gave the following example: ``If [it] can generate 10,000 lines of code and they are all correct\ldots you can rewrite most of PyTorch.'' Multiple participants referenced the continued exponential increase in task horizons (30 minutes, 1 hour, 40 hours) of the leading AI as quantifiable evidence of progress, citing a metric formalized by nonprofit METR around five months before the interviews were conducted\cite{kwa_measuring_2025}. This code captures participants that suggested long-horizon autonomous work represents a critical milestone, often citing METR. ``The most naive thing, I guess, is just the length, the duration of the work. And I think METR has an eval on this." (Participant 6) Participant 1 also mentioned the benchmark, saying ``If that's true, then you can put it on a calendar... when do we get to the 40-hour task horizon? Maybe we need another five doublings.'' 

\item \textbf{Existence Proofs} (occurred 5 times): Concrete demonstrations of autonomous research capabilities, even if not yet reliable or general. For example, success in mathematics competitions (IMO/IOI), novel algorithm discovery, or autonomous model training.

\item \textbf{Productivity Growth as Key Indicator} (occurred 3 times): Participant 6 argues that ASARA will look like ``increasing the labor supply'' and didn't see a sharp distinction. Multiple participants suggested we could identify ASARA as when ``researchers are not needed anymore.''

\item \textbf{Productivity Uplift} (occurred 3 times): Participant 25 suggested looking for when ASARA can ``provide 10x uplift to AI researchers.''

\item \textbf{Interpretability}: Loss of ability to understand model reasoning. Participant 6 said, ``I would say that one major milestone is if we can't understand how the model works at all.''

\item \textbf{Internal vs External Visibility}: Two participants mentioned AI companies deciding to keep their models internal for a competitive R\&D edge as a key milestone, before questions about whether participants expect internal deployments.

\end{itemize}

\subsection{Risk Perception and Concerns}\label{subsec:risk-perception}
\begin{figure}[h]
\centering
\includegraphics[width=\textwidth]{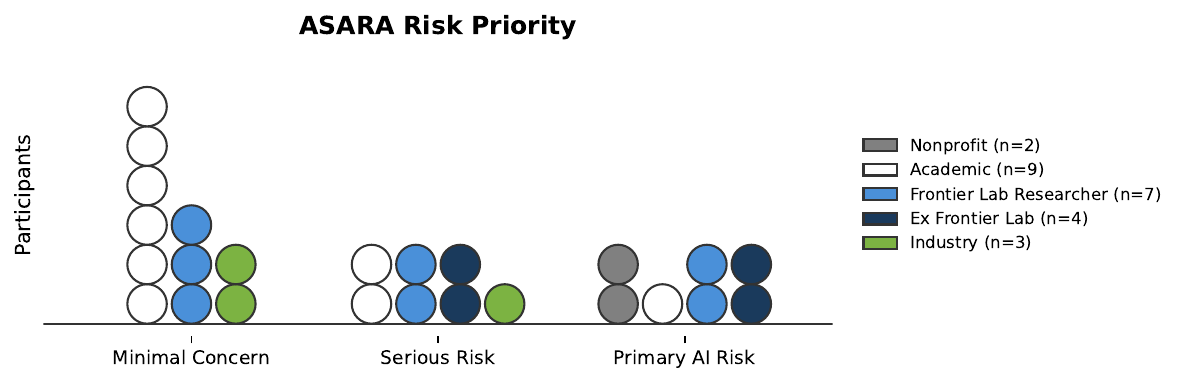}
\caption{Participant risk perception levels for ASARA, grouped by affiliation type.}
\label{fig:risk-perception}
\end{figure}

After establishing what participants envisioned for ASARA's development, the interviews turned to risk perception: whether participants view ASARA as a significant risk, and why. Figure \ref{fig:risk-perception} shows how many consider AI automating AI R\&D to be the primary driver of AI risk. Participants generally reported that their colleagues and organizations had similar views, suggesting that these perspectives reflect broader institutional positions.


Most of the participants viewed ASARA as carrying serious risks, with many identifying it as the main driver of AI risk. The most prevalent concerns that emerged included:

\begin{enumerate}
\item \textbf{Meta risk concern} (occurred in 18 transcripts): Participants viewed ASARA as amplifying all other AI risks. As Participant 2 explained: ``Recursive self-improvement just feels like it improves all capabilities across the board... it feels like, kind of like a superset worry relative to all the other ones.'' Participant 25 put it simply: ``It just speeds up other threat models.''

\item \textbf{Adaptation lag} (occurred in 17 transcripts): The fear that AI development would outpace human ability to understand or control it. Participants described scenarios where regulation, institutional responses, and human comprehension trail behind technological advancement.

\end{enumerate}

Some participants were worried about ASARA for other reasons:

\begin{enumerate}
\item \textbf{Power concentration} (occurred in 6 transcripts): Once an entity (company, nation, or group) gets ASARA, they can use it to accelerate their research faster than anyone else can catch up. Participant 8 was particularly concerned that ASARA creates a ``winner-take-all'' dynamic where the first to achieve it gains permanent control over the future of AI development.

\item \textbf{Loss of human relevance/employment} (occurred in 6 transcripts): Participant 12 says, ``Due to the recursive self-improvement, the companies don't need so many researchers... You become irrelevant.''

\item \textbf{Commercial incentives vs. safety} (occurred in 3 transcripts): Companies face pressure to use ASARA capabilities quickly for competitive advantage.
\end{enumerate}

Only two participants clearly dismissed the premise that AI systems could achieve recursive self-improvement. A larger subset of participants, largely academics, expressed minimal concern over ASARA risks without dismissing the possibility. Their reasoning was often similar to the binding constraints in Section \ref{subsubsec:bindingconstraints}. 16 transcripts include expressions of skepticism towards an explosive growth scenario due to technical constraints. However, some participants went beyond technical constraints. For example, Participant 6 argued that change would be gradual and observable: ``There's sort of an all or nothing view of it, as if it's something that's not happening today, but one day somebody will try it and get it working. But to me, it seems much more gradual."

\subsection{Views on Red Lines}\label{subsec:red_lines}
\begin{figure}[h]
\centering
\includegraphics[width=0.8\textwidth]{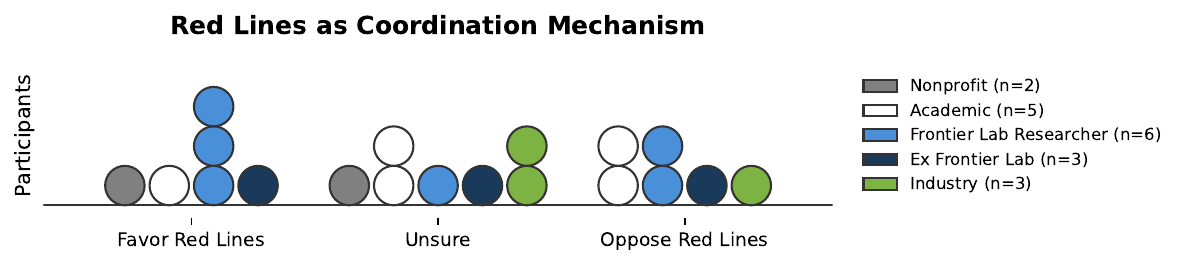}
\caption{Participant views on red lines for autonomous research grouped by affiliation type. Transcripts that did not express a clear perspective were not included.}
\label{fig:red_lines}
\end{figure}

Red lines are among the most aggressive proposals to AI risk mitigation: to trigger large responses from developers or governments at a particular threshold. Participants were divided on whether red lines represent an effective governance proposal, as seen in Figure \ref{fig:red_lines}. However, even supporters identified challenges with implementing this proposal.

\subsubsection{Implementation Challenges}
Across all positions, participants identified three primary challenges to making red lines effective in practice. First, participants identified a \textbf{specification problem}: the more precisely a threshold is defined (e.g., specific benchmark scores), the more it diverges from the actual risk it aims to capture. For instance, abstract thresholds like ``autonomous self-improvement'' prove difficult to operationalize and measure. Participant 24 explained this tension: ``The more concrete your red line, the more decoupled it becomes from the abstract intelligence explosion risk that you're worried about. And so there's just like an inherent struggle here.''

Second, the researchers highlighted \textbf{implementation challenges}, questioning the feasibility of verification and enforcement, one noted it would be ``a nightmare to get people to verify that they are adhering to these rules'' (P5, Frontier Lab Researcher). The definitional ambiguity of self-improvement, given ``any kind of AI includes some intelligence and a compiler optimization,'' complicates enforcement. Some pointed out that enforcement likely requires authoritarian levels of control or unjustifiable surveillance. 

Finally, participants were concerned about \textbf{timing and competitive dynamics}. On one hand, premature red lines could handicap genuinely beneficial development. One participant characterized red lines as ``the dumbest possible supervisor but the most trustworthy'' (P8, Former Frontier Lab Researcher), highlighting the trade-off between crude but transparent rules versus sophisticated oversight. Participant 10 explained the tension he saw as, ``Setting red lines too early could handicap beneficial development. But setting them too late means they're useless when you actually need them.''

\subsubsection{Transparency as a Possible Alternative}
Participants who were skeptical of red lines often preferred  transparency-based alternatives. Five participants clearly expressed preference for transparency-based approaches over rigid thresholds, often without being directly asked. Participant 25 argued for transparency over prohibition: ``I'd rather see mandatory reporting and monitoring than hard red lines... [so we can] maintain visibility into what's happening so we can make informed decisions as things develop.'' 

The appeal of transparency-based approaches by those skeptical of red lines can be partially explained by their uncertainty over AI's development trajectory. Participant 4 argued that fixed thresholds have too large an opportunity cost and might backfire: ``I think a dynamic governance proposal would be good... right now if we take any capability threshold in the near term, like AI being automated, we may get there and then it's just like not that big of a deal. And we may be actively harming ourselves actually by banning future progress.'' They also argued that any fixed threshold we set earlier would look ``pretty silly" in hindsight. Participant 11 also argued that a good policy must be adaptive: ``We need governance that can adapt as capabilities evolve. Fixed red lines are too rigid for a technology that's changing this fast.'' 


The transparency ideas most commonly proposed included mandatory reporting requirements for AI R\&D automation progress, increased visibility into internal deployments, and maintaining meaningful human oversight. 

\section{Discussion}\label{sec12}

\subsection{Schism Between Silicon Valley and Academia}

An epistemic divide emerged between frontier AI companies and academic institutions regarding ASARA development, reflecting deeper differences in organizational culture, incentives, and proximity to cutting-edge capabilities. All participants at frontier companies demonstrated active engagement with ASARA scenarios and reported regular internal discussions about recursive improvement dynamics. They note that these conversations are encouraged. For instance, Participant 6 said, ``I've been impressed so far, at least internally, with the discussions people have had that especially that the leaders are aware of these worries and talk openly about them, and bring them up on their own and encourage us to think about them." By contrast, academic participants often expressed limited consideration of these possibilities, with one noting: ``I don't even think the majority of AI researchers even think about this problem'' (P17, PhD Student). Participant 7 also brought this up when they said, ``When I ask questions like this, even if people in the labs are not believers in AGI or something, these are still discussions that we have at work. They're encouraged by the top... whereas, in an academic setting, it's like, you can't talk about this, or we'll, lose funding or something..."

This divide appears to be rooted in contrasting epistemic cultures. Academic participants characterized their institutional environment as skeptical, shaped by decades of unfulfilled AI promises and a peer review culture that rewards critical analysis. Participant 13 noted, ``skepticism is like a cultural thing in academia... if you're not criticizing, it means that you don't know enough.'' This skepticism extends to viewing intelligence explosion scenarios as professionally risky. Participant 18 notes that raising such themes risks being ``derided for being a bit of a crackpot'' in academic circles, despite these ideas circulating ``quite liberally in Silicon Valley.''

Conversely, participants from frontier companies described an environment oriented toward rapid capability development, with massive commercial incentives and firsthand experience with accelerating progress. Participant 10 described a sense of visceral evidence of exponential improvement via proximity: ``I think the large part is just having first-person experience of how fast things have gone... inside the labs, there's more of a sense of 'this is what it was like two years ago, and this is what it was two years before that, and here are all the arguments people brought up at that time, and they just turned out to be false.'" Participant 6 found it ``ironic that academics actually seem to be less clued in than many people in the general public or even politicians.'' Participants at labs also mentioned selection effects, wherein `deep believers' often join frontier labs from academia. More skeptical participants pointed out that frontier labs, who are beholden to investors rather than reviewers, are incentivized to over-promise and exaggerate their capabilities.

These differences extend beyond an ``optimism versus pessimism'' divide, reflecting divergences in how evidence is weighed, what risks are considered, what incentives are in place, and the futures that different groups imagine. 

\subsection{Cultural Variations in Perspectives}

Limited data makes international perspectives speculative, though the interviews suggest significant variation in how researchers conceptualize recursive AI improvement. Participants with exposure to Chinese AI research communities reported widespread skepticism about intelligence explosion scenarios. Participant 12, a professor from China, noted that ``when I talk about intelligence explosion in Weibo, even in China, people don't think it's possible... they think it's years ahead.'' They continue, saying that Chinese AI companies are ``very application oriented... they don't think about, oh, we want AI to take over the whole world. They just want to use AI to make some money.'' Multiple European perspectives, particularly from Germany and the United Kingdom, reflected similar skepticism compared to ``Silicon Valley's completely different mindset'' (P20, PhD Student).

\section{Limitations}
Methodologically, our interview sampling method was not random, which could affect sample representativeness and result validity. However, our sampling method purposefully sought to include diversity of perspectives among elite AI researchers. We paid deliberate attention to include contrarian views. My results are descriptive, targeted toward uncovering current views on a particular debate. The findings represent a snapshot of expert beliefs at a particular moment in AI’s development that are subject to change. Furthermore, the codes were developed inductively by the first author alone, so inter-rater reliability was not calculated. 

We do not make definitive claims about any consensus among AI researchers; our goal was to uncover qualitative explanations as to why participants expected particular outcomes, why mitigation might succeed or backfire, and why stakeholders might react in particular ways. 

\section{Conclusion}\label{sec13}

This study reveals divisions and commonalities within the AI research community surrounding AI research automation. Through 25 in-depth interviews with experts, we document converging technical visions despite disagreements about timelines, feasibility and governance approaches. 

The study revealed three key findings. First, an epistemic divide separates frontier AI companies from academia: researchers in the frontier AI labs engage with recursive improvement scenarios regularly with colleagues, and are encouraged by those above them. On the other hand, academic participants express systematic skepticism. I discuss how these perspectives may be shaped by proximity to cutting edge capabilities and opposing cultures. One implication from these findings is that policy discussions should include both of these groups.
Second, participants converged on ASARA as a high priority risk-model, largely because it is a ``meta risk'' which amplifies other risks by accelerating capabilities. Finally, researchers expressed skepticism about red lines as governance mechanisms, preferring transparency-based mitigation strategies.

\backmatter





\bmhead{Acknowledgements}

This research was funded by the ML Alignment \& Theory Scholars Program. We received helpful feedback from Raymond Douglas and Lily Stelling. 



\begin{appendices}
\section{Interview Protocol}\label{app:interview-protocol}

\subsection{Informed Consent Procedure}

Participants were read the following script before beginning the interview:

Thank you for agreeing to participate in this interview. My name is Severin Field, and I'm conducting this research supervised by professor David Krueger (Mila). We're exploring how AI might accelerate AI research itself, what this could look like, milestones we might see, and how different actors could respond.

Before we begin with questions, I will start recording to document your consent. So first, I want to remind you that: (1) You are free to skip any of the questions; (2) This interview will be recorded and transcribed within 48 hours. Transcripts are anonymized and recordings securely deleted, unless otherwise requested, so if you would like a copy you are welcome to have that. If your responses contain details that could indirectly identify you or others, they will be generalized. Anonymised transcripts will themselves be deleted once this research is complete; (3) Your participation is voluntary, so you may decline to answer, withdraw or adjust answers at any time before transcript anonymisation (within 48 hours). You are answering from your own perspective, i.e. not necessarily any organisation associated with you; (4) Your responses will be used for research purposes only and will be presented in aggregate form without any personally identifying details. Some de-identified quotes may be included in the final output, which will be linked to a participant ID linked to your occupation group, ie in this case [an X at a US/UK Y].

If you have any concerns, feel free to raise them now, or at any point during or after the interview. Do you have any questions before we begin? If you are comfortable proceeding, please confirm your consent to participate by saying `I confirm'.

\subsection{Interview Questions}

The semi-structured interview consisted of 14 core questions organized into three thematic sections, with conditional follow-up probes used as needed to elicit clarification or deeper responses.

\subsubsection{Opening}

\textbf{Q1.} What brought you to work on [your research]? What is your motivation?

\subsubsection{Section A: Conceptualizing ASARA}

\textbf{Q2.} How do you expect AI to accelerate AI R\&D in the coming years?

\textbf{Q3.} What do you think this will look like? \textit{(Follow-up probes included asking about LLMs vs. multi-system vs. new paradigm approaches, trajectory clarity, and confidence levels.)}

\textbf{Q4.} Have you read AI 2027? How does this view of AIs accelerating research compare to what you envision?\footnote{If participants were unfamiliar with AI 2027, the interviewer provided this description: The AI futures project recently released their AI 2027 scenario, which describes a rapid increase in AI capabilities from roughly AGI to super intelligence occurring during the year 2027, and driven primarily and then exclusively by ASARA. In their scenario, AI begins 2027 capable of performing research tasks, directed by human researchers, and progresses to autonomously executing open-ended research agendas using thousands of GPUs and open access to the Internet, before finally learning to autonomously improve itself far beyond human capabilities more quickly than humans can follow.}

\textbf{Q5.} How do you think Asara will contribute to or change the rate of AI progress? Are there ways you imagine it feeding back on itself in a positive feedback loop, for instance if AI systems can produce research that makes them better researchers?

\textbf{Q6.} Some thinkers expect an intelligence explosion.\footnote{Intelligence explosion was defined as: AI would quickly develop more advanced AI, which would itself develop even more advanced AI, but even faster and better than the last time, resulting in explosive growth in AI capabilities beyond our comprehension. Past a certain point, the process is self-sustaining and no longer requires human intervention.} Do you expect this dynamic? Are you skeptical of parts?

\textbf{Q7.} What are some key milestones that might indicate a rapid acceleration or paradigm shift in AI occurring due to ASARA? \textit{(Probes addressed observability, legibility, and concrete indicators of trajectory changes.)}

\subsubsection{Section B: Organizational Dynamics}

\textbf{Q8.} Suppose we develop AI systems that can significantly accelerate AI R\&D. Will organizations deploy these AI models? Keep internal? What will they be used for (e.g. making money, R\&D?)

\textbf{Q9.} Do you/your colleagues/team/other teams talk regularly about this? Are they worried? Excited? \textit{(Follow-ups explored whether such discussions were viewed as thought experiments versus serious organizational planning.)}

\subsubsection{Section C: Risk Assessment and Governance}

\textbf{Q10.} Are there any risks that you see arising from ASARA? Or ways in which you see ASARA significantly amplifying other risks? Why?

\textbf{Q11.} Compared to other AI risks, how worried are you about the risks associated with ASARA?

\textbf{Q12.} How do you think we could mitigate these risks? What mitigations are most promising? Could the mitigations cause more harm than good? How effective do you think they would be?

\textbf{Q13.} Are there any ``red lines'' you can think of wrt ASARA that we should not cross, or that should trigger a big response by AI developers or governments? What's your opinion on this?

\textbf{Q14.} \textit{(Optional)} Here's a basic proposal: international treaty where AIs should not be able to copy or improve themselves autonomously. Is this good/bad? What could go wrong? \textit{(When relevant, participants were asked about the IDIAS Beijing consensus, which includes ``No AI system should be able to copy or improve itself without explicit human approval and assistance.'')}

\subsubsection{Closing}

Interviews concluded with two open-ended questions: First, participants were asked if there was anything else they wanted to add or that we didn't cover that they thought was important. Second, they were asked to recommend other researchers who might have different perspectives or who had thought deeply about these topics.

Participants were reminded that all responses would be anonymized and that they could request a copy of the recording or transcript, adjust their answers, or have their transcript be made public/de-anonymised within 48 hours.

\section{Example Prompt for AI Assisted Coding}
\label{app:ai_prompts}
Below is an example of a prompt used. Similar structured prompts were adapted for other categorical variables to create figures. 

\begin{quotation}
``You are assisting with qualitative data extraction from anonymized interview transcripts. Your task is to identify and extract explicit statements related to whether participants expect AI labs to deploy ASARA models publicly or keep them internal.

\noindent\textbf{Input Variables:}
\begin{itemize}
\item Participant Number: [ID]
\item Organization Type: [Frontier Lab/Academia/Industry/Nonprofit/Ex-Frontier Lab]
\end{itemize}

\noindent\textbf{Research Question:}\\
``Do researchers expect AI labs to deploy ASARA models publicly or keep them internal? What reasoning do they provide?''

\noindent\textbf{Extraction Instructions:}
\begin{enumerate}
\item Identify all passages discussing deployment, transparency, or internal/external use
\item Extract verbatim quotes with context
\item Classify the participant's overall position using the following codes:
\begin{itemize}
\item 0 = No clear position expressed
\item 1 = Expects internal deployment
\item 2 = Uncertain/conditional
\item 3 = Expects public deployment
\end{itemize}
\end{enumerate}

\noindent\textbf{Output Format:}
\begin{itemize}
\item Categorical Code: [0-3]
\item Supporting Quotes: [List all relevant quotes]
\item Confidence Level: [Explicit statement / Implicit from context / Not addressed]"
\end{itemize}
\end{quotation}



\section{Participant-Suggested Milestones for ASARA Development}\label{app:milestones}

Researchers suggested various milestones that would indicate progress toward or away from rapid AI capability acceleration. This appendix presents key milestone categories identified through participant interviews.

\subsection{Task Horizons as a Core Observable Metric}

Participant 1 emphasized the importance of tracking task completion horizons:

\begin{quotation}
``Purely because we have some base rates for it, I like the METR task horizons indexing against 2024 onwards. If you look at their analysis, there appear to be two trends in the plot. Pre-2024, those numbers seem flaky—it's very hard to get accurately calibrated timings on very short tasks. From 2024 onwards, there appears to be a relatively clean curve, which corresponds to something like a doubling task horizon every four months at the 50\% threshold. For my own planning purposes, I'm just going to assume something like that is roughly true. If that's true, then you can put it on a calendar. An important benchmark they listed is when do we get to the 40-hour task horizon? Maybe we need another five doublings.'' (P1, Frontier Lab Researcher)
\end{quotation}

\subsection{Paradigm-Shifting Research Contributions}

Participant 17 articulated specific research capabilities that would constitute meaningful milestones:

\begin{quotation}
``If an AI researcher is able to propose a new idea for the next post-training paradigm and verify it actually works across, let's say, 30 billion scale, that would be a very concrete milestone saying AI research is making non-trivial contributions to empirical AI research. AI researchers coming up with better ways of doing pre-training that can speed up pre-training efficiency by significant margins—that's another example. I expect that to happen within the next one or two years.'' (P17, PhD Student, Stanford University)
\end{quotation}

They also noted domain-specific considerations:

\begin{quotation}
``There are domains like biology, chemistry, or drug discovery where in order to prove that you have found a positive solution—for example, a new drug that's effective—you have spent months or even years doing wet lab validation. For those cases, the feedback loop is a lot slower, and I expect the milestones to happen probably slower. But once it happens, it will be an even bigger milestone.'' (P17, PhD Student, Stanford University)
\end{quotation}

\subsection{Loss of Interpretability}

Participant 6 outlined a progression of interpretability-related milestones:

\begin{quotation}
``One major milestone is if we can't understand how the model works at all. A second milestone would be if it no longer works with a chain of thought, so maybe we understand how it works, but there's no interpretable reasoning happening. The third milestone would be if it's achieving feats in ways that are simply beyond our comprehension—even if it tried its very best to explain to us how it's working, it would fail in the same way that a 10-year-old simply wouldn't be able to understand how a language model today actually works.'' (P6, Frontier Lab Researcher)
\end{quotation}

\subsection{Existence Proofs and Productivity Measurements}

Participant 25 emphasized demonstrations over benchmarks:

\begin{quotation}
``Besides just the direct uplift, which is the actual thing that we all care about, existence proofs are really useful. Like `look, this AI system could train a 7B model entirely autonomously with just a lot of careful prompting and very little human intervention,' even if it can't do this very reliably, that's a meaningful sign—a very legible sign too.'' (P25, AI Researcher, Nonprofit)
\end{quotation}

They also suggested direct productivity measurement: ``The most useful thing to do would be to directly measure uplift at frontier companies that are trying to actually do AI research with AI systems.''

\subsection{Large-Scale Code Generation}

Participant 12 identified a concrete technical milestone:

\begin{quotation}
``If some model can generate 10,000 lines of code and they are all correct, that's one milestone, because if you can do that, you can rewrite most parts of PyTorch, and then you can have another version of PyTorch. Coding is a key milestone here.'' (P12, Professor, University in China)
\end{quotation}

\subsection{Internal Deployment Patterns}

Participant 24 highlighted organizational behavior as a signal:

\begin{quotation}
``If you're seeing a lot of internal deployment only for very advanced systems, [that would be an indicator of fast takeoff]'' (P24, Forecasting Researcher, Nonprofit)
\end{quotation}

\subsection{Evaluation and Judgment Capabilities}

Participant 21 identified evaluation capability as a critical bottleneck:

\begin{quotation}
``When there are benchmarks on understanding how machine learning models can evaluate a certain idea—when models start improving in that benchmark, especially if it's not trained on papers of that area but can generalize to a new area, and the benchmark numbers are improving—then maybe I would agree that we are going closer to the [intelligence explosion] scenario.'' (P21, AI Researcher, Big Tech Company)
\end{quotation}

\subsection{Competition Performance}

Participant 1 noted that performance on mathematics and coding competitions provides particularly compelling evidence: ``In the recent IMO and IOI competitions, they're basically the 99.99th percentile amongst coders or mathematicians.''

\subsection{Model Capability Uniformity}

Participant 23 introduced the concept of ``model spikiness'': the current reality where AI models excel in some domains (like mathematics) while being much weaker in others (like creative writing). They suggested that decreasing spikiness might indicate increased likelihood of rapid capability acceleration, reasoning that uniform capability across domains would enable true recursive improvement, as models could improve themselves along all necessary dimensions simultaneously rather than getting stuck in domains where they lack competence.
\end{appendices}

\bibliography{sn-bibliography}


\begin{thebibliography}{23}
\ifx \bisbn   \undefined \def \bisbn  #1{ISBN #1}\fi
\ifx \binits  \undefined \def \binits#1{#1}\fi
\ifx \bauthor  \undefined \def \bauthor#1{#1}\fi
\ifx \batitle  \undefined \def \batitle#1{#1}\fi
\ifx \bjtitle  \undefined \def \bjtitle#1{#1}\fi
\ifx \bvolume  \undefined \def \bvolume#1{\textbf{#1}}\fi
\ifx \byear  \undefined \def \byear#1{#1}\fi
\ifx \bissue  \undefined \def \bissue#1{#1}\fi
\ifx \bfpage  \undefined \def \bfpage#1{#1}\fi
\ifx \blpage  \undefined \def \blpage #1{#1}\fi
\ifx \burl  \undefined \def \burl#1{\textsf{#1}}\fi
\ifx \doiurl  \undefined \def \doiurl#1{\url{https://doi.org/#1}}\fi
\ifx \betal  \undefined \def \betal{\textit{et al.}}\fi
\ifx \binstitute  \undefined \def \binstitute#1{#1}\fi
\ifx \binstitutionaled  \undefined \def \binstitutionaled#1{#1}\fi
\ifx \bctitle  \undefined \def \bctitle#1{#1}\fi
\ifx \beditor  \undefined \def \beditor#1{#1}\fi
\ifx \bpublisher  \undefined \def \bpublisher#1{#1}\fi
\ifx \bbtitle  \undefined \def \bbtitle#1{#1}\fi
\ifx \bedition  \undefined \def \bedition#1{#1}\fi
\ifx \bseriesno  \undefined \def \bseriesno#1{#1}\fi
\ifx \blocation  \undefined \def \blocation#1{#1}\fi
\ifx \bsertitle  \undefined \def \bsertitle#1{#1}\fi
\ifx \bsnm \undefined \def \bsnm#1{#1}\fi
\ifx \bsuffix \undefined \def \bsuffix#1{#1}\fi
\ifx \bparticle \undefined \def \bparticle#1{#1}\fi
\ifx \barticle \undefined \def \barticle#1{#1}\fi
\bibcommenthead
\ifx \bconfdate \undefined \def \bconfdate #1{#1}\fi
\ifx \botherref \undefined \def \botherref #1{#1}\fi
\ifx \url \undefined \def \url#1{\textsf{#1}}\fi
\ifx \bchapter \undefined \def \bchapter#1{#1}\fi
\ifx \bbook \undefined \def \bbook#1{#1}\fi
\ifx \bcomment \undefined \def \bcomment#1{#1}\fi
\ifx \oauthor \undefined \def \oauthor#1{#1}\fi
\ifx \citeauthoryear \undefined \def \citeauthoryear#1{#1}\fi
\ifx \endbibitem  \undefined \def \endbibitem {}\fi
\ifx \bconflocation  \undefined \def \bconflocation#1{#1}\fi
\ifx \arxivurl  \undefined \def \arxivurl#1{\textsf{#1}}\fi
\csname PreBibitemsHook\endcsname

\bibitem[\protect\citeauthoryear{Wiseman and McClements}{2025}]{wiseman_how_2025}
\begin{botherref}
\oauthor{\bsnm{Wiseman}, \binits{J.}},
\oauthor{\bsnm{McClements}, \binits{D.}}:
How much economic growth from {AI} should we expect, how soon?
(2025).
\url{https://inferencemagazine.substack.com/p/how-much-economic-growth-from-ai}
Accessed 2025-10-29
\end{botherref}
\endbibitem

\bibitem[\protect\citeauthoryear{Good}{1966}]{alt_speculations_1966}
\begin{botherref}
\oauthor{\bsnm{Good}, \binits{I.J.}}:
Speculations {Concerning} the {First} {Ultraintelligent} {Machine}**{Based} on talks given in a {Conference} on the {Conceptual} {Aspects} of {Biocommunications}, {Neuropsychiatric} {Institute}, {University} of {California}, {Los} {Angeles}, {October} 1962; and in the {Artificial} {Intelligence} {Sessions} of the {Winter} {General} {Meetings} of the {IEEE}, {January} 1963 [1, 46].{The} first draft of this monograph was completed in {April} 1963, and the present slightly amended version in {May} 1964.{I} am much indebted to {Mrs}. {Euthie} {Anthony} of {IDA} for the arduous task of typing.
Advances in {Computers},
vol. 6,
pp. 31--88.
Elsevier
(1966).
\doiurl{10.1016/S0065-2458(08)60418-0} .
ISSN: 0065-2458.
\url{https://www.sciencedirect.com/science/article/pii/S0065245808604180}
\end{botherref}
\endbibitem

\bibitem[\protect\citeauthoryear{Yudkowsky}{}]{yudkowsky_intelligence_nodate}
\begin{botherref}
\oauthor{\bsnm{Yudkowsky}, \binits{E.}}:
Intelligence {Explosion} {Microeconomics}
\end{botherref}
\endbibitem

\bibitem[\protect\citeauthoryear{Castelvecchi}{2025}]{castelvecchi_deepmind_2025}
\begin{barticle}
\bauthor{\bsnm{Castelvecchi}, \binits{D.}}:
\batitle{{DeepMind} and {OpenAI} models solve maths problems at level of top students}.
\bjtitle{Nature}
\bvolume{644}(\bissue{8075}),
\bfpage{20}--\blpage{20}
(\byear{2025})
\doiurl{10.1038/d41586-025-02343-x} .
Accessed 2025-08-31
\end{barticle}
\endbibitem

\bibitem[\protect\citeauthoryear{{OpenAI}}{2025}]{openai_agi_2025}
\begin{botherref}
\oauthor{\bsnm{{OpenAI}}}:
{AGI} progress, surprising breakthroughs, and the road ahead --- the {OpenAI} {Podcast} {Ep}. 5
(2025).
\url{https://www.youtube.com/watch?v=yBzStBK6Z8c}
Accessed 2025-08-31
\end{botherref}
\endbibitem

\bibitem[\protect\citeauthoryear{Chollet}{2018}]{chollet_implausibility_2018}
\begin{botherref}
\oauthor{\bsnm{Chollet}, \binits{F.}}:
The implausibility of intelligence explosion.
Publication Title: Medium
(2018).
\url{https://medium.com/@francois.chollet/the-impossibility-of-intelligence-explosion-5be4a9eda6ec}
Accessed 2025-08-31
\end{botherref}
\endbibitem

\bibitem[\protect\citeauthoryear{Hausenloy}{2025}]{hausenloy_red_2025}
\begin{botherref}
\oauthor{\bsnm{Hausenloy}, \binits{J.}}:
Red {Lines} for {Recursive} {Self}-{Improvement}
(2025).
\url{https://firstscattering.com/p/red-lines-for-recursive-self-improvement}
Accessed 2025-08-31
\end{botherref}
\endbibitem

\bibitem[\protect\citeauthoryear{Zoumpalova}{2025}]{zoumpalova_global_2025}
\begin{botherref}
\oauthor{\bsnm{Zoumpalova}, \binits{T.}}:
Global {Red} {Lines} for {AI}: {A} {Three}-{Part} {Series}.
Publication Title: The Future Society
(2025).
\url{https://thefuturesociety.org/airedlines/}
Accessed 2025-08-31
\end{botherref}
\endbibitem

\bibitem[\protect\citeauthoryear{}{}]{noauthor_idais-beijing_nodate}
\begin{botherref}
{IDAIS}-{Beijing}.
Publication Title: International Dialogues on AI Safety.
\url{https://idais.ai/dialogue/idais-beijing/}
Accessed 2025-08-31
\end{botherref}
\endbibitem

\bibitem[\protect\citeauthoryear{Eth}{}]{eth_will_nodate}
\begin{botherref}
\oauthor{\bsnm{Eth}, \binits{D.}}:
Will {AI} {R}\&{D} {Automation} {Cause} a {Software} {Intelligence} {Explosion}?
Publication Title: Forethought.
\url{https://www.forethought.org/research/will-ai-r-and-d-automation-cause-a-software-intelligence-explosion}
Accessed 2025-08-31
\end{botherref}
\endbibitem

\bibitem[\protect\citeauthoryear{Bostrom}{2014}]{bostrom_superintelligence_2014}
\begin{bbook}
\bauthor{\bsnm{Bostrom}, \binits{N.}}:
\bbtitle{Superintelligence: {Paths}, Dangers, Strategies}.
\bsertitle{Superintelligence: {Paths}, dangers, strategies}.
\bpublisher{Oxford University Press},
\blocation{New York, NY, US}
(\byear{2014})
\end{bbook}
\endbibitem

\bibitem[\protect\citeauthoryear{Starace et~al.}{2025}]{starace_paperbench_2025}
\begin{botherref}
\oauthor{\bsnm{Starace}, \binits{G.}},
\oauthor{\bsnm{Jaffe}, \binits{O.}},
\oauthor{\bsnm{Sherburn}, \binits{D.}},
\oauthor{\bsnm{Aung}, \binits{J.}},
\oauthor{\bsnm{Chan}, \binits{J.S.}},
\oauthor{\bsnm{Maksin}, \binits{L.}},
\oauthor{\bsnm{Dias}, \binits{R.}},
\oauthor{\bsnm{Mays}, \binits{E.}},
\oauthor{\bsnm{Kinsella}, \binits{B.}},
\oauthor{\bsnm{Thompson}, \binits{W.}},
\oauthor{\bsnm{Heidecke}, \binits{J.}},
\oauthor{\bsnm{Glaese}, \binits{A.}},
\oauthor{\bsnm{Patwardhan}, \binits{T.}}:
{PaperBench}: {Evaluating} {AI}'s {Ability} to {Replicate} {AI} {Research}.
arXiv
(2025).
\doiurl{10.48550/arXiv.2504.01848} .
\url{http://arxiv.org/abs/2504.01848}
Accessed 2025-08-31
\end{botherref}
\endbibitem

\bibitem[\protect\citeauthoryear{Wijk et~al.}{2025}]{wijk_re-bench_2025}
\begin{botherref}
\oauthor{\bsnm{Wijk}, \binits{H.}},
\oauthor{\bsnm{Lin}, \binits{T.}},
\oauthor{\bsnm{Becker}, \binits{J.}},
\oauthor{\bsnm{Jawhar}, \binits{S.}},
\oauthor{\bsnm{Parikh}, \binits{N.}},
\oauthor{\bsnm{Broadley}, \binits{T.}},
\oauthor{\bsnm{Chan}, \binits{L.}},
\oauthor{\bsnm{Chen}, \binits{M.}},
\oauthor{\bsnm{Clymer}, \binits{J.}},
\oauthor{\bsnm{Dhyani}, \binits{J.}},
\oauthor{\bsnm{Ericheva}, \binits{E.}},
\oauthor{\bsnm{Garcia}, \binits{K.}},
\oauthor{\bsnm{Goodrich}, \binits{B.}},
\oauthor{\bsnm{Jurkovic}, \binits{N.}},
\oauthor{\bsnm{Karnofsky}, \binits{H.}},
\oauthor{\bsnm{Kinniment}, \binits{M.}},
\oauthor{\bsnm{Lajko}, \binits{A.}},
\oauthor{\bsnm{Nix}, \binits{S.}},
\oauthor{\bsnm{Sato}, \binits{L.}},
\oauthor{\bsnm{Saunders}, \binits{W.}},
\oauthor{\bsnm{Taran}, \binits{M.}},
\oauthor{\bsnm{West}, \binits{B.}},
\oauthor{\bsnm{Barnes}, \binits{E.}}:
{RE}-{Bench}: {Evaluating} frontier {AI} {R}\&{D} capabilities of language model agents against human experts.
arXiv
(2025).
\doiurl{10.48550/arXiv.2411.15114} .
\url{http://arxiv.org/abs/2411.15114}
Accessed 2025-08-31
\end{botherref}
\endbibitem

\bibitem[\protect\citeauthoryear{Chan et~al.}{2025}]{chan_mle-bench_2025}
\begin{botherref}
\oauthor{\bsnm{Chan}, \binits{J.S.}},
\oauthor{\bsnm{Chowdhury}, \binits{N.}},
\oauthor{\bsnm{Jaffe}, \binits{O.}},
\oauthor{\bsnm{Aung}, \binits{J.}},
\oauthor{\bsnm{Sherburn}, \binits{D.}},
\oauthor{\bsnm{Mays}, \binits{E.}},
\oauthor{\bsnm{Starace}, \binits{G.}},
\oauthor{\bsnm{Liu}, \binits{K.}},
\oauthor{\bsnm{Maksin}, \binits{L.}},
\oauthor{\bsnm{Patwardhan}, \binits{T.}},
\oauthor{\bsnm{Weng}, \binits{L.}},
\oauthor{\bsnm{M{\k{a}}dry}, \binits{A.}}:
{MLE}-bench: {Evaluating} {Machine} {Learning} {Agents} on {Machine} {Learning} {Engineering}.
arXiv
(2025).
\doiurl{10.48550/arXiv.2410.07095} .
\url{http://arxiv.org/abs/2410.07095}
Accessed 2025-08-31
\end{botherref}
\endbibitem

\bibitem[\protect\citeauthoryear{Yang et~al.}{2025}]{yang_rd-agent_2025}
\begin{botherref}
\oauthor{\bsnm{Yang}, \binits{X.}},
\oauthor{\bsnm{Yang}, \binits{X.}},
\oauthor{\bsnm{Fang}, \binits{S.}},
\oauthor{\bsnm{Zhang}, \binits{Y.}},
\oauthor{\bsnm{Wang}, \binits{J.}},
\oauthor{\bsnm{Xian}, \binits{B.}},
\oauthor{\bsnm{Li}, \binits{Q.}},
\oauthor{\bsnm{Li}, \binits{J.}},
\oauthor{\bsnm{Xu}, \binits{M.}},
\oauthor{\bsnm{Li}, \binits{Y.}},
\oauthor{\bsnm{Pan}, \binits{H.}},
\oauthor{\bsnm{Zhang}, \binits{Y.}},
\oauthor{\bsnm{Liu}, \binits{W.}},
\oauthor{\bsnm{Shen}, \binits{Y.}},
\oauthor{\bsnm{Chen}, \binits{W.}},
\oauthor{\bsnm{Bian}, \binits{J.}}:
R\&{D}-{Agent}: {An} {LLM}-{Agent} {Framework} {Towards} {Autonomous} {Data} {Science}
(2025).
Accessed 2025-11-26
\end{botherref}
\endbibitem

\bibitem[\protect\citeauthoryear{Kwa et~al.}{2025}]{kwa_measuring_2025}
\begin{botherref}
\oauthor{\bsnm{Kwa}, \binits{T.}},
\oauthor{\bsnm{West}, \binits{B.}},
\oauthor{\bsnm{Becker}, \binits{J.}},
\oauthor{\bsnm{Deng}, \binits{A.}},
\oauthor{\bsnm{Garcia}, \binits{K.}},
\oauthor{\bsnm{Hasin}, \binits{M.}},
\oauthor{\bsnm{Jawhar}, \binits{S.}},
\oauthor{\bsnm{Kinniment}, \binits{M.}},
\oauthor{\bsnm{Rush}, \binits{N.}},
\oauthor{\bsnm{Arx}, \binits{S.V.}},
\oauthor{\bsnm{Bloom}, \binits{R.}},
\oauthor{\bsnm{Broadley}, \binits{T.}},
\oauthor{\bsnm{Du}, \binits{H.}},
\oauthor{\bsnm{Goodrich}, \binits{B.}},
\oauthor{\bsnm{Jurkovic}, \binits{N.}},
\oauthor{\bsnm{Miles}, \binits{L.H.}},
\oauthor{\bsnm{Nix}, \binits{S.}},
\oauthor{\bsnm{Lin}, \binits{T.}},
\oauthor{\bsnm{Parikh}, \binits{N.}},
\oauthor{\bsnm{Rein}, \binits{D.}},
\oauthor{\bsnm{Sato}, \binits{L.J.K.}},
\oauthor{\bsnm{Wijk}, \binits{H.}},
\oauthor{\bsnm{Ziegler}, \binits{D.M.}},
\oauthor{\bsnm{Barnes}, \binits{E.}},
\oauthor{\bsnm{Chan}, \binits{L.}}:
Measuring {AI} {Ability} to {Complete} {Long} {Tasks}.
arXiv
(2025).
\doiurl{10.48550/arXiv.2503.14499} .
\url{http://arxiv.org/abs/2503.14499}
Accessed 2025-08-31
\end{botherref}
\endbibitem

\bibitem[\protect\citeauthoryear{Becker et~al.}{2025}]{becker_measuring_2025}
\begin{botherref}
\oauthor{\bsnm{Becker}, \binits{J.}},
\oauthor{\bsnm{Rush}, \binits{N.}},
\oauthor{\bsnm{Barnes}, \binits{E.}},
\oauthor{\bsnm{Rein}, \binits{D.}}:
Measuring the {Impact} of {Early}-2025 {AI} on {Experienced} {Open}-{Source} {Developer} {Productivity}.
arXiv
(2025).
\doiurl{10.48550/arXiv.2507.09089} .
\url{http://arxiv.org/abs/2507.09089}
Accessed 2025-08-31
\end{botherref}
\endbibitem

\bibitem[\protect\citeauthoryear{Owen}{2024}]{owen_interviewing_2024}
\begin{botherref}
\oauthor{\bsnm{Owen}, \binits{D.}}:
Interviewing {AI} researchers on automation of {AI} {R}\&{D}.
Publication Title: Epoch AI
(2024).
\url{https://epoch.ai/blog/interviewing-ai-researchers-on-automation-of-ai-rnd}
Accessed 2025-08-31
\end{botherref}
\endbibitem

\bibitem[\protect\citeauthoryear{Leibowich}{}]{leibowich_could_nodate}
\begin{botherref}
\oauthor{\bsnm{Leibowich}, \binits{J.}}:
Could {Advanced} {AI} {Accelerate} the {Pace} of {AI} {Progress}? {Interviews} with {AI} {Researchers} by {Jared} {Leibowich}, {Nikola} {Jurkovic}, {Tom} {Davidson} :: {SSRN}.
\url{https://papers.ssrn.com/sol3/papers.cfm?abstract_id=5115692}
Accessed 2025-08-31
\end{botherref}
\endbibitem

\bibitem[\protect\citeauthoryear{{Davidson}}{}]{davidson_how_nodate}
\begin{botherref}
\oauthor{\bsnm{{Davidson}}}:
How quick and big would a software intelligence explosion be?
Publication Title: Forethought.
\url{https://www.forethought.org/research/how-quick-and-big-would-a-software-intelligence-explosion-be}
Accessed 2025-08-31
\end{botherref}
\endbibitem

\bibitem[\protect\citeauthoryear{Kokotajlo}{}]{kokotajlo_ai_nodate}
\begin{botherref}
\oauthor{\bsnm{Kokotajlo}, \binits{D.}}:
{AI} 2027.
\url{https://ai-2027.com/}
Accessed 2025-08-31
\end{botherref}
\endbibitem

\bibitem[\protect\citeauthoryear{Saunders}{}]{saunders_anonymising_nodate}
\begin{botherref}
\oauthor{\bsnm{Saunders}, \binits{B.}}:
Anonymising interview data: challenges and compromise in practice - {Benjamin} {Saunders}, {Jenny} {Kitzinger}, {Celia} {Kitzinger}, 2015.
\url{https://journals.sagepub.com/doi/full/10.1177/1468794114550439}
Accessed 2025-08-31
\end{botherref}
\endbibitem

\bibitem[\protect\citeauthoryear{{OpenAI} et~al.}{2024}]{openai_gpt-4_2024}
\begin{botherref}
\oauthor{\bsnm{{OpenAI}}},
\oauthor{\bsnm{Achiam}, \binits{J.}},
\oauthor{\bsnm{Adler}, \binits{S.}},
\oauthor{\bsnm{Agarwal}, \binits{S.}},
\oauthor{\bsnm{Ahmad}, \binits{L.}},
\oauthor{\bsnm{Akkaya}, \binits{I.}},
\oauthor{\bsnm{Aleman}, \binits{F.L.}},
\oauthor{\bsnm{Almeida}, \binits{D.}},
\oauthor{\bsnm{Altenschmidt}, \binits{J.}},
\oauthor{\bsnm{Altman}, \binits{S.}},
\oauthor{\bsnm{Anadkat}, \binits{S.}},
\oauthor{\bsnm{Avila}, \binits{R.}},
\oauthor{\bsnm{Babuschkin}, \binits{I.}},
\oauthor{\bsnm{Balaji}, \binits{S.}},
\oauthor{\bsnm{Balcom}, \binits{V.}},
\oauthor{\bsnm{Baltescu}, \binits{P.}},
\oauthor{\bsnm{Bao}, \binits{H.}},
\oauthor{\bsnm{Bavarian}, \binits{M.}},
\oauthor{\bsnm{Belgum}, \binits{J.}},
\oauthor{\bsnm{Bello}, \binits{I.}},
\oauthor{\bsnm{Berdine}, \binits{J.}},
\oauthor{\bsnm{Bernadett-Shapiro}, \binits{G.}},
\oauthor{\bsnm{Berner}, \binits{C.}},
\oauthor{\bsnm{Bogdonoff}, \binits{L.}},
\oauthor{\bsnm{Boiko}, \binits{O.}},
\oauthor{\bsnm{Boyd}, \binits{M.}},
\oauthor{\bsnm{Brakman}, \binits{A.-L.}},
\oauthor{\bsnm{Brockman}, \binits{G.}},
\oauthor{\bsnm{Brooks}, \binits{T.}},
\oauthor{\bsnm{Brundage}, \binits{M.}},
\oauthor{\bsnm{Button}, \binits{K.}},
\oauthor{\bsnm{Cai}, \binits{T.}},
\oauthor{\bsnm{Campbell}, \binits{R.}},
\oauthor{\bsnm{Cann}, \binits{A.}},
\oauthor{\bsnm{Carey}, \binits{B.}},
\oauthor{\bsnm{Carlson}, \binits{C.}},
\oauthor{\bsnm{Carmichael}, \binits{R.}},
\oauthor{\bsnm{Chan}, \binits{B.}},
\oauthor{\bsnm{Chang}, \binits{C.}},
\oauthor{\bsnm{Chantzis}, \binits{F.}},
\oauthor{\bsnm{Chen}, \binits{D.}},
\oauthor{\bsnm{Chen}, \binits{S.}},
\oauthor{\bsnm{Chen}, \binits{R.}},
\oauthor{\bsnm{Chen}, \binits{J.}},
\oauthor{\bsnm{Chen}, \binits{M.}},
\oauthor{\bsnm{Chess}, \binits{B.}},
\oauthor{\bsnm{Cho}, \binits{C.}},
\oauthor{\bsnm{Chu}, \binits{C.}},
\oauthor{\bsnm{Chung}, \binits{H.W.}},
\oauthor{\bsnm{Cummings}, \binits{D.}},
\oauthor{\bsnm{Currier}, \binits{J.}},
\oauthor{\bsnm{Dai}, \binits{Y.}},
\oauthor{\bsnm{Decareaux}, \binits{C.}},
\oauthor{\bsnm{Degry}, \binits{T.}},
\oauthor{\bsnm{Deutsch}, \binits{N.}},
\oauthor{\bsnm{Deville}, \binits{D.}},
\oauthor{\bsnm{Dhar}, \binits{A.}},
\oauthor{\bsnm{Dohan}, \binits{D.}},
\oauthor{\bsnm{Dowling}, \binits{S.}},
\oauthor{\bsnm{Dunning}, \binits{S.}},
\oauthor{\bsnm{Ecoffet}, \binits{A.}},
\oauthor{\bsnm{Eleti}, \binits{A.}},
\oauthor{\bsnm{Eloundou}, \binits{T.}},
\oauthor{\bsnm{Farhi}, \binits{D.}},
\oauthor{\bsnm{Fedus}, \binits{L.}},
\oauthor{\bsnm{Felix}, \binits{N.}},
\oauthor{\bsnm{Fishman}, \binits{S.P.}},
\oauthor{\bsnm{Forte}, \binits{J.}},
\oauthor{\bsnm{Fulford}, \binits{I.}},
\oauthor{\bsnm{Gao}, \binits{L.}},
\oauthor{\bsnm{Georges}, \binits{E.}},
\oauthor{\bsnm{Gibson}, \binits{C.}},
\oauthor{\bsnm{Goel}, \binits{V.}},
\oauthor{\bsnm{Gogineni}, \binits{T.}},
\oauthor{\bsnm{Goh}, \binits{G.}},
\oauthor{\bsnm{Gontijo-Lopes}, \binits{R.}},
\oauthor{\bsnm{Gordon}, \binits{J.}},
\oauthor{\bsnm{Grafstein}, \binits{M.}},
\oauthor{\bsnm{Gray}, \binits{S.}},
\oauthor{\bsnm{Greene}, \binits{R.}},
\oauthor{\bsnm{Gross}, \binits{J.}},
\oauthor{\bsnm{Gu}, \binits{S.S.}},
\oauthor{\bsnm{Guo}, \binits{Y.}},
\oauthor{\bsnm{Hallacy}, \binits{C.}},
\oauthor{\bsnm{Han}, \binits{J.}},
\oauthor{\bsnm{Harris}, \binits{J.}},
\oauthor{\bsnm{He}, \binits{Y.}},
\oauthor{\bsnm{Heaton}, \binits{M.}},
\oauthor{\bsnm{Heidecke}, \binits{J.}},
\oauthor{\bsnm{Hesse}, \binits{C.}},
\oauthor{\bsnm{Hickey}, \binits{A.}},
\oauthor{\bsnm{Hickey}, \binits{W.}},
\oauthor{\bsnm{Hoeschele}, \binits{P.}},
\oauthor{\bsnm{Houghton}, \binits{B.}},
\oauthor{\bsnm{Hsu}, \binits{K.}},
\oauthor{\bsnm{Hu}, \binits{S.}},
\oauthor{\bsnm{Hu}, \binits{X.}},
\oauthor{\bsnm{Huizinga}, \binits{J.}},
\oauthor{\bsnm{Jain}, \binits{S.}},
\oauthor{\bsnm{Jain}, \binits{S.}},
\oauthor{\bsnm{Jang}, \binits{J.}},
\oauthor{\bsnm{Jiang}, \binits{A.}},
\oauthor{\bsnm{Jiang}, \binits{R.}},
\oauthor{\bsnm{Jin}, \binits{H.}},
\oauthor{\bsnm{Jin}, \binits{D.}},
\oauthor{\bsnm{Jomoto}, \binits{S.}},
\oauthor{\bsnm{Jonn}, \binits{B.}},
\oauthor{\bsnm{Jun}, \binits{H.}},
\oauthor{\bsnm{Kaftan}, \binits{T.}},
\oauthor{\bsnm{Kaiser}, \binits{{\L}.}},
\oauthor{\bsnm{Kamali}, \binits{A.}},
\oauthor{\bsnm{Kanitscheider}, \binits{I.}},
\oauthor{\bsnm{Keskar}, \binits{N.S.}},
\oauthor{\bsnm{Khan}, \binits{T.}},
\oauthor{\bsnm{Kilpatrick}, \binits{L.}},
\oauthor{\bsnm{Kim}, \binits{J.W.}},
\oauthor{\bsnm{Kim}, \binits{C.}},
\oauthor{\bsnm{Kim}, \binits{Y.}},
\oauthor{\bsnm{Kirchner}, \binits{J.H.}},
\oauthor{\bsnm{Kiros}, \binits{J.}},
\oauthor{\bsnm{Knight}, \binits{M.}},
\oauthor{\bsnm{Kokotajlo}, \binits{D.}},
\oauthor{\bsnm{Kondraciuk}, \binits{{\L}.}},
\oauthor{\bsnm{Kondrich}, \binits{A.}},
\oauthor{\bsnm{Konstantinidis}, \binits{A.}},
\oauthor{\bsnm{Kosic}, \binits{K.}},
\oauthor{\bsnm{Krueger}, \binits{G.}},
\oauthor{\bsnm{Kuo}, \binits{V.}},
\oauthor{\bsnm{Lampe}, \binits{M.}},
\oauthor{\bsnm{Lan}, \binits{I.}},
\oauthor{\bsnm{Lee}, \binits{T.}},
\oauthor{\bsnm{Leike}, \binits{J.}},
\oauthor{\bsnm{Leung}, \binits{J.}},
\oauthor{\bsnm{Levy}, \binits{D.}},
\oauthor{\bsnm{Li}, \binits{C.M.}},
\oauthor{\bsnm{Lim}, \binits{R.}},
\oauthor{\bsnm{Lin}, \binits{M.}},
\oauthor{\bsnm{Lin}, \binits{S.}},
\oauthor{\bsnm{Litwin}, \binits{M.}},
\oauthor{\bsnm{Lopez}, \binits{T.}},
\oauthor{\bsnm{Lowe}, \binits{R.}},
\oauthor{\bsnm{Lue}, \binits{P.}},
\oauthor{\bsnm{Makanju}, \binits{A.}},
\oauthor{\bsnm{Malfacini}, \binits{K.}},
\oauthor{\bsnm{Manning}, \binits{S.}},
\oauthor{\bsnm{Markov}, \binits{T.}},
\oauthor{\bsnm{Markovski}, \binits{Y.}},
\oauthor{\bsnm{Martin}, \binits{B.}},
\oauthor{\bsnm{Mayer}, \binits{K.}},
\oauthor{\bsnm{Mayne}, \binits{A.}},
\oauthor{\bsnm{McGrew}, \binits{B.}},
\oauthor{\bsnm{McKinney}, \binits{S.M.}},
\oauthor{\bsnm{McLeavey}, \binits{C.}},
\oauthor{\bsnm{McMillan}, \binits{P.}},
\oauthor{\bsnm{McNeil}, \binits{J.}},
\oauthor{\bsnm{Medina}, \binits{D.}},
\oauthor{\bsnm{Mehta}, \binits{A.}},
\oauthor{\bsnm{Menick}, \binits{J.}},
\oauthor{\bsnm{Metz}, \binits{L.}},
\oauthor{\bsnm{Mishchenko}, \binits{A.}},
\oauthor{\bsnm{Mishkin}, \binits{P.}},
\oauthor{\bsnm{Monaco}, \binits{V.}},
\oauthor{\bsnm{Morikawa}, \binits{E.}},
\oauthor{\bsnm{Mossing}, \binits{D.}},
\oauthor{\bsnm{Mu}, \binits{T.}},
\oauthor{\bsnm{Murati}, \binits{M.}},
\oauthor{\bsnm{Murk}, \binits{O.}},
\oauthor{\bsnm{M{\'e}ly}, \binits{D.}},
\oauthor{\bsnm{Nair}, \binits{A.}},
\oauthor{\bsnm{Nakano}, \binits{R.}},
\oauthor{\bsnm{Nayak}, \binits{R.}},
\oauthor{\bsnm{Neelakantan}, \binits{A.}},
\oauthor{\bsnm{Ngo}, \binits{R.}},
\oauthor{\bsnm{Noh}, \binits{H.}},
\oauthor{\bsnm{Ouyang}, \binits{L.}},
\oauthor{\bsnm{O'Keefe}, \binits{C.}},
\oauthor{\bsnm{Pachocki}, \binits{J.}},
\oauthor{\bsnm{Paino}, \binits{A.}},
\oauthor{\bsnm{Palermo}, \binits{J.}},
\oauthor{\bsnm{Pantuliano}, \binits{A.}},
\oauthor{\bsnm{Parascandolo}, \binits{G.}},
\oauthor{\bsnm{Parish}, \binits{J.}},
\oauthor{\bsnm{Parparita}, \binits{E.}},
\oauthor{\bsnm{Passos}, \binits{A.}},
\oauthor{\bsnm{Pavlov}, \binits{M.}},
\oauthor{\bsnm{Peng}, \binits{A.}},
\oauthor{\bsnm{Perelman}, \binits{A.}},
\oauthor{\bsnm{Peres}, \binits{F.d.A.B.}},
\oauthor{\bsnm{Petrov}, \binits{M.}},
\oauthor{\bsnm{Pinto}, \binits{H.P.d.O.}},
\oauthor{\bsnm{{Michael}}},
\oauthor{\bsnm{{Pokorny}}},
\oauthor{\bsnm{Pokrass}, \binits{M.}},
\oauthor{\bsnm{Pong}, \binits{V.H.}},
\oauthor{\bsnm{Powell}, \binits{T.}},
\oauthor{\bsnm{Power}, \binits{A.}},
\oauthor{\bsnm{Power}, \binits{B.}},
\oauthor{\bsnm{Proehl}, \binits{E.}},
\oauthor{\bsnm{Puri}, \binits{R.}},
\oauthor{\bsnm{Radford}, \binits{A.}},
\oauthor{\bsnm{Rae}, \binits{J.}},
\oauthor{\bsnm{Ramesh}, \binits{A.}},
\oauthor{\bsnm{Raymond}, \binits{C.}},
\oauthor{\bsnm{Real}, \binits{F.}},
\oauthor{\bsnm{Rimbach}, \binits{K.}},
\oauthor{\bsnm{Ross}, \binits{C.}},
\oauthor{\bsnm{Rotsted}, \binits{B.}},
\oauthor{\bsnm{Roussez}, \binits{H.}},
\oauthor{\bsnm{Ryder}, \binits{N.}},
\oauthor{\bsnm{Saltarelli}, \binits{M.}},
\oauthor{\bsnm{Sanders}, \binits{T.}},
\oauthor{\bsnm{Santurkar}, \binits{S.}},
\oauthor{\bsnm{Sastry}, \binits{G.}},
\oauthor{\bsnm{Schmidt}, \binits{H.}},
\oauthor{\bsnm{Schnurr}, \binits{D.}},
\oauthor{\bsnm{Schulman}, \binits{J.}},
\oauthor{\bsnm{Selsam}, \binits{D.}},
\oauthor{\bsnm{Sheppard}, \binits{K.}},
\oauthor{\bsnm{Sherbakov}, \binits{T.}},
\oauthor{\bsnm{Shieh}, \binits{J.}},
\oauthor{\bsnm{Shoker}, \binits{S.}},
\oauthor{\bsnm{Shyam}, \binits{P.}},
\oauthor{\bsnm{Sidor}, \binits{S.}},
\oauthor{\bsnm{Sigler}, \binits{E.}},
\oauthor{\bsnm{Simens}, \binits{M.}},
\oauthor{\bsnm{Sitkin}, \binits{J.}},
\oauthor{\bsnm{Slama}, \binits{K.}},
\oauthor{\bsnm{Sohl}, \binits{I.}},
\oauthor{\bsnm{Sokolowsky}, \binits{B.}},
\oauthor{\bsnm{Song}, \binits{Y.}},
\oauthor{\bsnm{Staudacher}, \binits{N.}},
\oauthor{\bsnm{Such}, \binits{F.P.}},
\oauthor{\bsnm{Summers}, \binits{N.}},
\oauthor{\bsnm{Sutskever}, \binits{I.}},
\oauthor{\bsnm{Tang}, \binits{J.}},
\oauthor{\bsnm{Tezak}, \binits{N.}},
\oauthor{\bsnm{Thompson}, \binits{M.B.}},
\oauthor{\bsnm{Tillet}, \binits{P.}},
\oauthor{\bsnm{Tootoonchian}, \binits{A.}},
\oauthor{\bsnm{Tseng}, \binits{E.}},
\oauthor{\bsnm{Tuggle}, \binits{P.}},
\oauthor{\bsnm{Turley}, \binits{N.}},
\oauthor{\bsnm{Tworek}, \binits{J.}},
\oauthor{\bsnm{Uribe}, \binits{J.F.C.}},
\oauthor{\bsnm{Vallone}, \binits{A.}},
\oauthor{\bsnm{Vijayvergiya}, \binits{A.}},
\oauthor{\bsnm{Voss}, \binits{C.}},
\oauthor{\bsnm{Wainwright}, \binits{C.}},
\oauthor{\bsnm{Wang}, \binits{J.J.}},
\oauthor{\bsnm{Wang}, \binits{A.}},
\oauthor{\bsnm{Wang}, \binits{B.}},
\oauthor{\bsnm{Ward}, \binits{J.}},
\oauthor{\bsnm{Wei}, \binits{J.}},
\oauthor{\bsnm{Weinmann}, \binits{C.J.}},
\oauthor{\bsnm{Welihinda}, \binits{A.}},
\oauthor{\bsnm{Welinder}, \binits{P.}},
\oauthor{\bsnm{Weng}, \binits{J.}},
\oauthor{\bsnm{Weng}, \binits{L.}},
\oauthor{\bsnm{Wiethoff}, \binits{M.}},
\oauthor{\bsnm{Willner}, \binits{D.}},
\oauthor{\bsnm{Winter}, \binits{C.}},
\oauthor{\bsnm{Wolrich}, \binits{S.}},
\oauthor{\bsnm{Wong}, \binits{H.}},
\oauthor{\bsnm{Workman}, \binits{L.}},
\oauthor{\bsnm{Wu}, \binits{S.}},
\oauthor{\bsnm{Wu}, \binits{J.}},
\oauthor{\bsnm{Wu}, \binits{M.}},
\oauthor{\bsnm{Xiao}, \binits{K.}},
\oauthor{\bsnm{Xu}, \binits{T.}},
\oauthor{\bsnm{Yoo}, \binits{S.}},
\oauthor{\bsnm{Yu}, \binits{K.}},
\oauthor{\bsnm{Yuan}, \binits{Q.}},
\oauthor{\bsnm{Zaremba}, \binits{W.}},
\oauthor{\bsnm{Zellers}, \binits{R.}},
\oauthor{\bsnm{Zhang}, \binits{C.}},
\oauthor{\bsnm{Zhang}, \binits{M.}},
\oauthor{\bsnm{Zhao}, \binits{S.}},
\oauthor{\bsnm{Zheng}, \binits{T.}},
\oauthor{\bsnm{Zhuang}, \binits{J.}},
\oauthor{\bsnm{Zhuk}, \binits{W.}},
\oauthor{\bsnm{Zoph}, \binits{B.}}:
{GPT}-4 {Technical} {Report}.
arXiv
(2024).
\doiurl{10.48550/arXiv.2303.08774} .
\url{http://arxiv.org/abs/2303.08774}
Accessed 2025-08-31
\end{botherref}
\endbibitem

\end{thebibliography}

\end{document}